# Fractional charges in emergent neutral modes at the integer quantum Hall effect

Hiroyuki Inoue[1], Anna Grivnin[1], Nissim Ofek[2], Izhar Neder[3], Moty Heiblum[1], Vladimir Umansky[1] and Diana Mahalu[1]

[1]Braun Center for Submicron Research, Department of Condensed Matter Physics, Weizmann Institute of Science, Rehovot, Israel

[2]Departments of Physics and Applied Physics, Yale University, New Haven, CT, USA

[3]Raymond and Beverly Sackler School of Physics and Astronomy, Tel-Aviv University, Tel Aviv, 69978, Israel

**Charge fractionalization is a possible emergent excitation in a low-dimensional system of interacting electrons. A known example is that of fractional charges in the fractional quantum Hall effect (FQHE) regime, which is a consequence of strong Coulomb interaction among the electrons whose kinetic energy is quenched by the strong magnetic field. Alternatively, the integer QHE (IQHE), with electrons behaving largely as** *independent particles* **in Landau levels (LLs), lacks such fractionalization. However, for integer LLs filling $v=2, 3,...$, electrons propagate in copropagating adjacent chiral edge channels, and thus interact and modify the non-interacting LLs. For example, at $v=2$, an electron injected selectively into a single non-interacting (bare) edge channel is expected to decompose into a 'fast' mode and a 'slow' mode in the region of interaction; each mode carry fractional charges shared between the two bare channels. Here, we report our sensitive shot noise measurement that affirms the presence of such fractionalization in $v=2$. Injecting partitioned current into a 'hot' edge channel led to low frequency shot noise in the adjacent currentless 'cold' edge channel after it had been partitioned. Controlling the partitioning of the hot and cold channels allowed a determination of the fractional charges in both channels as well as the channels' velocity difference. This approach can be extended to study interaction in two-dimensional systems with a topology dictating edge channels transport.**

Charge fractionalization is an exotic manifestation of low dimensional correlated electrons. Fractionally charged quasiparticles in the FQHE[1-6] and in quantum wires[7] provide examples. Here, we employ a high mobility two dimensional electron gas in the IQHE regime and at filling factor $v$=2 (spin split lowest LL). At low-energies the transport takes place solely at the sample's periphery via gapless chiral (downstream) edge channels, with channel conductance $G_0=e^2/h$, where $e$ the electron charge and $h$ the Planck constant[8-9]. Due to the spatial separation between the channels, each channel can be manipulated (reflected, transmitted, or redirected) individually; customarily with a quantum point contact (QPCs) constriction. A recently discovered upstream neutral edge modes in hole-conjugate FQHE states, resulting from edge reconstruction[10-12], revealed the importance of inter-channel Coulomb interaction (in this case it was also accompanied by inter channel tunneling). Evidently, Coulomb interaction takes place also in the IQHE regime; as had been already reported and analyzed in studies that observed 'lobe structures' in the visibility of interferometers[13-14]; interaction mediated dephasing[15-18]; and energy equilibration among edge channels[19-23]. Moreover, inter-channel interaction between copropagating channels had been already predicted to give rise to downstream charge density modes propagating at different velocities and carry fractional excitations[24-29]. While the existence of modes with different dispersion relations had been already reported[30], the charge of their excitations has not been determined yet. Here, we demonstrate an observation of fractionally charged excitations in the IQHE, with two interacting chiral edge channels, via low frequency shot noise measurements[3,31-33].

Employing a channel selective quantum point contact (QPC) allows injecting partitioned electrons in one of the two copropagating channels (a 'hot' channel). Inter-channel interaction (without tunneling) will induce local charge imbalance in the 'cold' channel, which upon partitioning by another QPC, will bare a finite currentless shot noise (fluctuations around zero net current). Measuring the dependence of that noise on the partitioned 'hot' current, allowed the determination the interaction strength and the excitations charge.

Before delving into the details of our experiment, it might be illuminating to present an intuitive model of the system. The resulting excitations in the two channels (1 & 2) can be regarded as fractionally charged dipoles flowing downstream, with the spin degree of freedom play no role here. Assume short range inter-channel interaction $uh\delta(x_1-x_2)$, the energy density of the interacting channels $\varepsilon = h(v_1\rho_1^2 + v_2\rho_2^2 + u\rho_1\rho_2)$, with $\rho_1(x_1)$ and $\rho_2(x_2)$ being the number densities and $\upsilon_1$ and $\upsilon_2$ the velocities of the non-interacting channels. As shown in Fig. 1a, injecting an electron in channel 1 (via a partitioning QPC1 with probability $T_1$), gives birth to two modes distributed between channels 1 and 2: (*i*) a 'slow' mode, consisting of a particle-like charged $+(1-\alpha)e$ and a hole-like charged $-\beta e$, in channel 1 & 2, respectively; (*ii*) a 'fast' mode, consisting of particles-like charged $+\alpha e$ and $+\beta e$, in channel 1 & 2, respectively  Here, $\alpha=(1+\cos\theta)/2$, $\beta=0.5\sin\theta$, and *tan θ* $=u/\Delta\upsilon$ ($0<\theta<\pi/2$) with $\Delta\upsilon=\upsilon_1-\upsilon_2$. The two modes propagate at velocities $v_\pm = \bar{v} \pm \frac{1}{2}\sqrt{(\Delta v)^2 + u^2}$, where $\bar{v}=(\upsilon_1+\upsilon_2)/2$ and + (-) stand for fast (slow). Noting that when the two channels have equal velocities, $\alpha=\beta$, and the 'slow' mode is neutral. Moreover, for $\upsilon_->0$, $4\upsilon_1\upsilon_2>u^2$ must be satisfied. Measuring the charges of the fractionalized wave packets by fast chopping in order to separate the 'fast' mode from the 'slow' one, necessitates rather challenging high frequency measurements. Hence, since channel 2 (the 'cold' channel) carries always zero net current but a fluctuating 'neutral excitation', $+\beta e$ and $-\beta e$, we chose to characterize these fractionally charged quasiparticles via continuous low frequency shot noise measurements, which arise by partitioning stochastically the stream of quasiparticles in channel 2.

## Results

**Measurement scheme**

Our 2DEG was embedded in a ubiquitous GaAs-AlGaAs heterostructure. The resultant patterning is shown in the SEM micrograph in Fig. 1b with a magnified view of the core part below. The light blue region is a mesa where the 2DEG exists; the light gray curves are metallic gate electrodes; yellow pads are ohmic contacts, where S1 & S2 denote source contacts, G grounds, and A1 & A2 amplifier contacts (each loaded with a resonant circuit, $f_0\sim790$kHz, followed by a cryogenic amplifier). The two QPC constrictions, separated by an interaction region, $l=8\mu$m, with its potential being modified by an additional side gate off the mesa, SG, whose voltage was kept at -500mV (gates are green in Figs. 1c & d). In configuration C1 the outer channel plays the role of the 'hot' channel (channel 1), while in configuration C2 the inner channel plays that role. In configuration C1 (Fig. 1c), the source current $2I$ (thick red lines) is shared equally between the two edge channels. Constriction QPC1 fully reflects the inner channel while partitions the outer channel ($T_1$, $R_1$, dotted heavy red line). Two 'cold' edge channels, emanating from the G contacts, also impinge at QPC1 (thin blue lines), with the outer is fully transmitted and the inner fully reflected. The reflected 'cold' channel (thick blue line) flows in close proximity to the partitioned outer channel, with both reaching QPC2. There, the outer channel is fully transmitted and inner one is being partitioned ($T_2$, $R_2$, dotted thick blue line), with its excess current noise (spectral density, $S_i$) monitored at A1. We employ configuration C1 (C2, shown in Fig. 1d, with the role of the two channels reversed) to measure the dependence of excess noise of the cold channel on $T_1$ of the 'hot' channel ($T_2$ of the 'cold' channel). Such arrangement is chosen since the transmission probability $T_1$ of the outer 'hot' channel is fairly constant with energy. We stress that suppressing the inter-channel tunneling current (below $5 \times 10^{-4}$ of source current) was crucial for reliable results.

The measured noise is composed of excess shot noise, thermal (Johnson-Nyquist) noise[31], and preamplifier current and voltage noises[32]. The 'hot' channel is being partitioned stochastically at QPC1 with transmission probability $T_1$, with a resultant 'white' excess noise, $S_i = 2eIT_1(1-T_1)(\coth x - x^{-1})$, where $S_i$ its 'zero frequency' spectral density (frequency<< $eV/h$), with $x = eIG_0^{-1}/2k_B\Theta$, $k_B$ Boltzmann constant[31-33]. The spectral density depends linearly (quadratically) on high (low) $I$ for a given electron temperature, which was found to be $\Theta$~20mK. The spectral density of the partitioned 'hot' channel was measured as function of $T_1$. Its normalized magnitude was plotted in the inset of Fig. 2a. The expected dependence $\propto T_1(1-T_1)$ is observed. Note that, due to the multi-terminal configuration, the amplifier is fed by a constant Hall resistance, and thus the current noise of the preamplifier and the Nyquist noise were both independent of QPC transmission. However, in the present experiment we expect to measure shot noise without net current in the partitioned, unbiased, 'cold', channel - a peculiar situation.

**The fractionalization noise**

The net current and excess noise $S_i$ in A1 were measured in configuration C1 at current $I$=-1.6 to +0.4nA, for different QPC transmissions $T_1$ (0.06-0.94), while QPC2 was kept at constant transmission $T_2$=0.5 (Fig. 2a). As the injected current (in absolute value) increased, the excess noise $S_i$ increased; however, without net current. The dependence of the excess noise in the partitioned 'cold' channel on $I$ resemble roughly a standard excess noise. However, it is interesting to notice that the low noise rounding (near zero current) is wider than the corresponding one in the hot channel. We return to this point later again. A dependence of the excess noise on $T_1$ is shown in Fig. 2b (normalized to $T_1 \cong 0.5$); obeying a simple dependence $[T_1(1-T_1)]^{\gamma_1}$, with $\gamma_1 = 0.70$ (for comparison, curves with $\gamma_1 = 0.5$ & $1.0$ are also plotted).

Similar measurements were repeated with configuration C2, where the role of the two channels was reversed. The dependence of $S_i$ in the outer channel (now the 'cold' channel) on the injected current into the inner channel (now the 'hot' channel, partitioned with $T_1 = 0.5$) is plotted in Fig. 3a. Different partitioning $T_2$ of the 'cold' channel (in range 0.1–1) were employed. In the same manner as previously, a dependence $[T_2(1-T_2)]^{\gamma_2}$ of the noise was found (Fig. 3b); with $\gamma_2 = 0.95$. This time the partitioning appears to be nearly binomial in $T_2$.

Since the predicted fractional excitations in the 'hot' channel is $\alpha e$ and $(1-\alpha)e$, it is only natural to ask whether those quasiparticles can be measured via shot noise. Partitioning the 'hot' channel with QPC2, after it interacted with the 'cold' channel, led, however, to the 'boring' spectral density of independent partitioned electrons, $S_i = 2eIT_1T_2(1-T_1T_2)$, for a wide range of $T_1$ and $T_2$ (data not shown); namely, revealing charge $e$. This result can be understood by realizing that the low frequency shot noise can only reflect tunneling events of electrons in the QPC; being in the IQHE regime, only electrons are allowed to back scatter by the QPC.

## Discussion

**Comparing with theory**

Several recent theories considered our present experimental scheme[27-29]. Particularly, Ref. 29 provided a platform for how to extract the *mixing angle* $\theta$ (*tan* $\theta = u/\Delta \upsilon$) and the fractional charge $\beta e$ ($\beta = 0.5 sin\ \theta$) from the measured spectral density $S_i$ of the 'cold' channel at zero temperature. This paper provides the missing connection between $\theta$ and the strength of the noise (expressed by the Fano factor $F$ (see Fig. 4a)) as well as with the noise dependence on $T_1$, namely, $\gamma_1$ (see Fig. 4b). Defining the Fano factor $F(\theta) = S_i/S_{Ref}$, which reflects the charge of the partitioned quasiparticles $e^* = Fe$, where the reference spectral density is $S_{Ref} = 4eIT_1R_1R_2$, being the excess noise due to stochastically back scattering of a random train of electrons and holes. The prediction assumes only inter-channel interaction, namely, void of interaction with an external environment and for $R_2=1-T_2\ll1$. With $0<\theta<\pi/2$, $\gamma_1$ was calculated to span $0.68\leq\gamma_1(\theta)\leq1$. As shown in Fig. 4b, for $\theta=\pi/2$ the calculated $\gamma_1=0.68$ - the case when the bare channels' velocities are equal, $\Delta\upsilon=0$, and $\beta=\alpha=0.5$ (the 'slow' mode, shared by the two LL's, is *neutral*). Approaching the non-interacting case, $\theta\to0$, the 'cold' channel has a diminishing noise, and $\gamma_1\to1$. Note also that $\gamma_1=0.5$ (outside the scale in Fig. 4b) stands for the two channels fully thermalizing along the interaction region due to interaction with the environment. Plotting in Fig. 4a the expected fractionalization charge $\beta=(sin\ \theta)/2$, which was evaluated with the simple model above, we find a nice agreement with the numerical evaluation of expected fractional charge expressed by $F(\theta)$[29].

**The fractional charge and velocities**

Noting that obtaining $\gamma_2=0.95$, with excess noise being nearly binomial in $T_2$, namely, linearly dependent on $R_2$ for $R_2\to0$, justifies the perturbative treatment in $R_2$ in the theory[29]. Comparing our data with Ref. 29, our determined $\gamma_1=0.70$ falls within the predicted range (void of interaction with the environment), leading to a mixing angle $\theta\sim\pi/3.1$, $u/\Delta\upsilon=1.56$, and $F(\pi/3.1)=0.47$ (Fig. 4). This prediction is compared with the measured Fano factor (the slope of the excess noise in the range $-1.2nA\leq I \leq-0.8nA$ for $T_1=0.5$ and $T_2=0.1$ divided by with $4eT_1R_1T_2R_2$); found remarkably to be $F=0.46$. Verifying consistently the fractional charges are $\beta e=0.42e$ in the 'cold' channel and $\alpha e=0.77e$ in the 'hot' channel.

While most of the parameters of the system had been extracted, the strength of the interaction $u$ is still missing. In a similar configuration (performed by our group in Ref. 15); applying a DC bias of 19μV to the inner channel of $\nu=2$, resulted in $2\pi$ phase shift in a coupled interferometer formed by the inner channel being 10μm long; suggesting an addition of one electron. Therefore, the mutual capacitance between the channels can be estimated as $C\sim0.8fF/\mu m$. Note that the logarithmic dependent Coulomb interaction distance makes the exact number less important. The conversion relation to $u=e^2/hC$ leads to $u=4.5\times10^4$m/s, yielding $\Delta\upsilon=2.9\times10^4$m/s (with the minimum average velocity $2.7\times10^4$m/s, deduced from $\upsilon_-=0$). The excessive 'rounding' of the excess noise *vs.* current traces may result from an overlap of the fractional wave packets ($\pm\beta e$) at low $I$; thus suppressing the measured noise in the 'cold' channel. However, additional experiments must be performed to verify this effect.

With all this said, it might be useful to provide also an intuitive picture of the mechanism leading to the excess noise in the 'cold' channel. Partitioning the DC current in the 'hot' channel by QPC1 leads to a wideband current fluctuations with a cutoff at frequency $I/e$. Obviously, the inter-channel capacitance $Cl=e^2l/hu$ induces high frequency displacement current noise in the 'cold' channel, with a low frequency cutoff that depends on the $Cl/G_0$ time constant (much higher than our measurement frequency). However, stochastic partitioning of the unbiased channel by QPC2 redistributes the high frequency spectrum over the entire spectrum (up to the cutoff frequency) - yet with zero net current.

In summary, observing neutral modes, with zero net current, in an unbiased, interacting, edge state in the IQHE regime (filling factor, $v=2$), allowed a determination of the fractional excitations that form the neutral modes. The neutral modes where characterized by an emerging shot noise after partitioning by a quantum point contact; allowed also the determination of the interaction energy and the relative velocities of the two channels void of interaction. Our scheme opens a way to probe Coulomb correlations in multiple 1D channels of other QHEs and topological insulators.

## Methods

### Experimental setup

The device was fabricated on a GaAs/AlGaAs heterostructure; with 2DEG embedded 130nm below the surface whose carrier density is $8.2\times10^{10}\,\text{cm}^{-2}$ and dark mobility is $4.2\times10^{6}\,\text{cm}^{2}/\text{Vs}$ at 4.2K. The constrictions QPC1 and QPC2, formed by negatively biased split-gates (5nm Ti/15nm Au) with a 600nm wide opening, are separated by center-to-center distance $l$=8μm. Contacts S1, S2, A1, A2, and G, are made of the ubiquitous alloyed AuGeNi. The grounded contacts were tied directly to the cold finger of the dilution refrigerator at 10mK. All the measurements were done at the magnetic field $B$=1.7T, where the plateau of the bulk filling factor 2 with longitudinal resistance $R_{xx}$ ~ 0Ω and Hall resistance $R_{xy}=(2G_0)^{-1}$~12.9 kΩ. The noise signal at A1 and A2, filtered with an LC circuit tuned to 790 kHz, was first amplified by a cooled, home-made, preamplifier with voltage gain 11.6, and subsequently by a room temperature amplifier (NF-220F5) with voltage gain 200, followed by a spectrum analyzer with the bandwidth of 10 kHz. The total background noise was 280pV/√Hz at the resonant frequency. For the transmission measurement, 0.5μV$_{RMS}$ at the resonant frequency, with or without an accompanying DC voltage, was applied at source S1 (S2) and measured at A1 (A2) with the bandwidth of 30Hz. The tunneling current was monitored by increasing the excitation amplitude up to ten times on top on the DC current biases.

## Acknowledgements

We thank B. Rosenow, M. Milletarì, Y. Oreg, and G. Viola for helpful discussions and H. K. Choi for his technical help. We acknowledge the partial support of the Israeli Science Foundation (ISF), the Minerva foundation, the US-Israel Bi-National Science Foundation (BSF), and the European Research Council under the European Community's Seventh Framework Program (FP7/2007-2013)/ERC, Grant agreement # 227716.

**Figures**

**Figure 1. Schematics of the experiment. a**, An ordered train of electrons driven from the source (S) transmit the QPC1 with a probability $T_1$ and decompose into fast and slow modes. The fast mode consists of fractional charges $\alpha e$ $(\beta e)$ on channel 1 (2) and the slow mode consists of fractional charges $(1-\alpha)e, (-\beta e)$ on channel 1 (2). The pairs of $\pm \beta e$ are partitioned at QPC2 with a transmission probability $T_2$, generating a low-frequency shot noise to be detected at the amplifier (A). **b**, SEM image of the employed device fabricated on GaAs/AlGaAs. The mesa is blue-highlighted. The yellow pads are ohmic contacts, sources (S1&S2), amplifiers (A1&A2), grounds (G). The gray curves are metallic gates, QPC1, QPC2, and side gate (SG). Below, a magnified view of the core part. The region between the QPCs is the interaction region. **c**, The configuration 1 (C1). S1 and A1 were employed. The red and the blue arrows are the hot and the cold channels. The biased outer channel is the channel 1 here. QPC1 and QPC2 were tuned to transmission probabilities $T_1$ and $T_2$ for corresponding channels. For various $T_1$ transmitting only the biased outer channel, $T_2$ was set to 0.5 to reflect the unbiased inner channel to A1. **d,** Configuration 2 (C2). S2 and A2 were employed. The biased inner channel is the channel 1 here. $T_1$ was set to 0.5 and the reflected inner channel was directed to QPC2 with various $T_2$ to partition the fluctuating but unbiased outer channel.

**Figure 2. | Excess noise as a function of $T_1$. a**, The excess noise traces, measured with the configuration C1, as a function of $I$ for several $T_1$ with fixed $T_2$=0.5 are shown. Noise traces of selected $T_1$ are shown. In the inset plots, by scaling the trace at $T_1 = 0.5$ to unity, relative noise magnitude of the hot channel at different $T_1$ behaves as the independent Fermionic one. **b**, Relative magnitude of the excess noise as a function of $T_1$, normalized to the one at $T_1$=0.5. The excess noise is proportional to $[T_1(1-T_1)]^{\gamma_1}$, where $\gamma_1$=0.70 for $T_2$=0.5. For comparison, curves with $\gamma_1$=0.5 and 1.0 are also plotted.

**Figure 3. | Excess noise as a function of $T_2$. a**, The excess noise traces, measured with the configuration C2, as a function of $I$ for selected $T_2$ are shown. Similar excess noise was again observed. **b**, Relative magnitude of the excess noise as a function of $T_2$, normalized to the one at $T_2$=0.5, is plotted. The excess noise is proportional to $[T_2(1-T_2)]^{\gamma_2}$, where $\gamma_2$=0.95 with $T_1$=0.5.

**Figure 4 | Theoretical Fano factor $F(\theta)$ and exponent $\gamma_1(\theta)$** (Ref. 29). Theoretical plots relevant to the present setup and the fractionalization excess noise $S_i$. The mixing angle $\theta$ therein was modified to fit our notations. **a**, The Fano factor $F=S_i/S_{Ref}$ ($S_{Ref}=4eIT_1R_1R_2$) representing the fractional charge in the cold channel ($e^*=Fe$), plotted as function of the mixing angle. The red dots are the theory and the blue curve depicts $\beta=0.5\sin\theta$, based on the simple model in the paper. The latter model and the numerical one[29] show a remarkable agreement. **b**, The exponent $\gamma_1$ plotted as function of the mixing angle based on the numerical computation[29]. The experimentally obtained $\gamma_1$ yields a mixing angle $\theta\sim\pi/3.1$ ($\tan\theta=u/\Delta\upsilon=1.56$), which reads $F=0.47$, $\beta=0.42$ and $\alpha=0.77$.

Figure 1

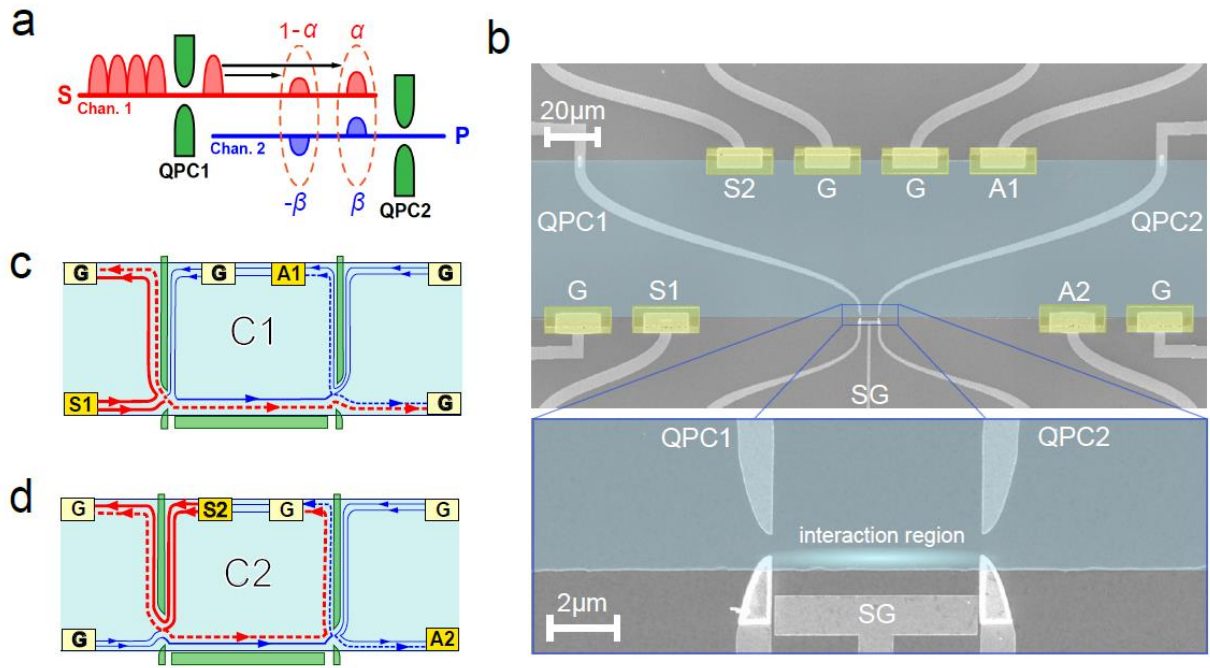

Figure 2

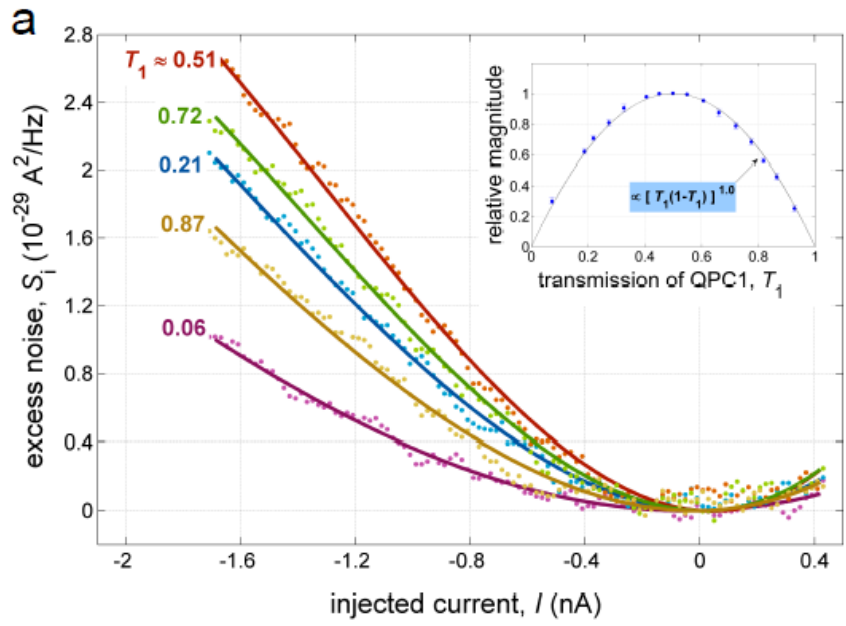

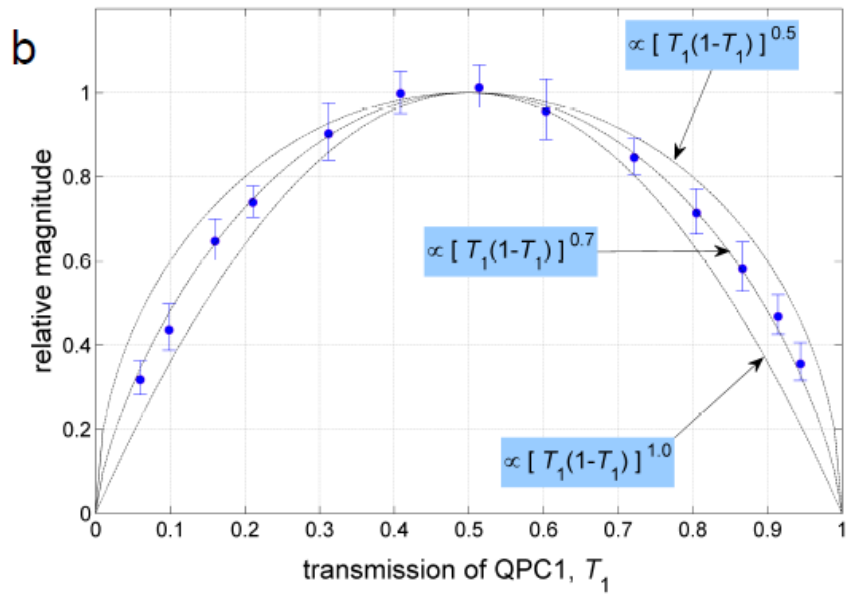

Figure 3

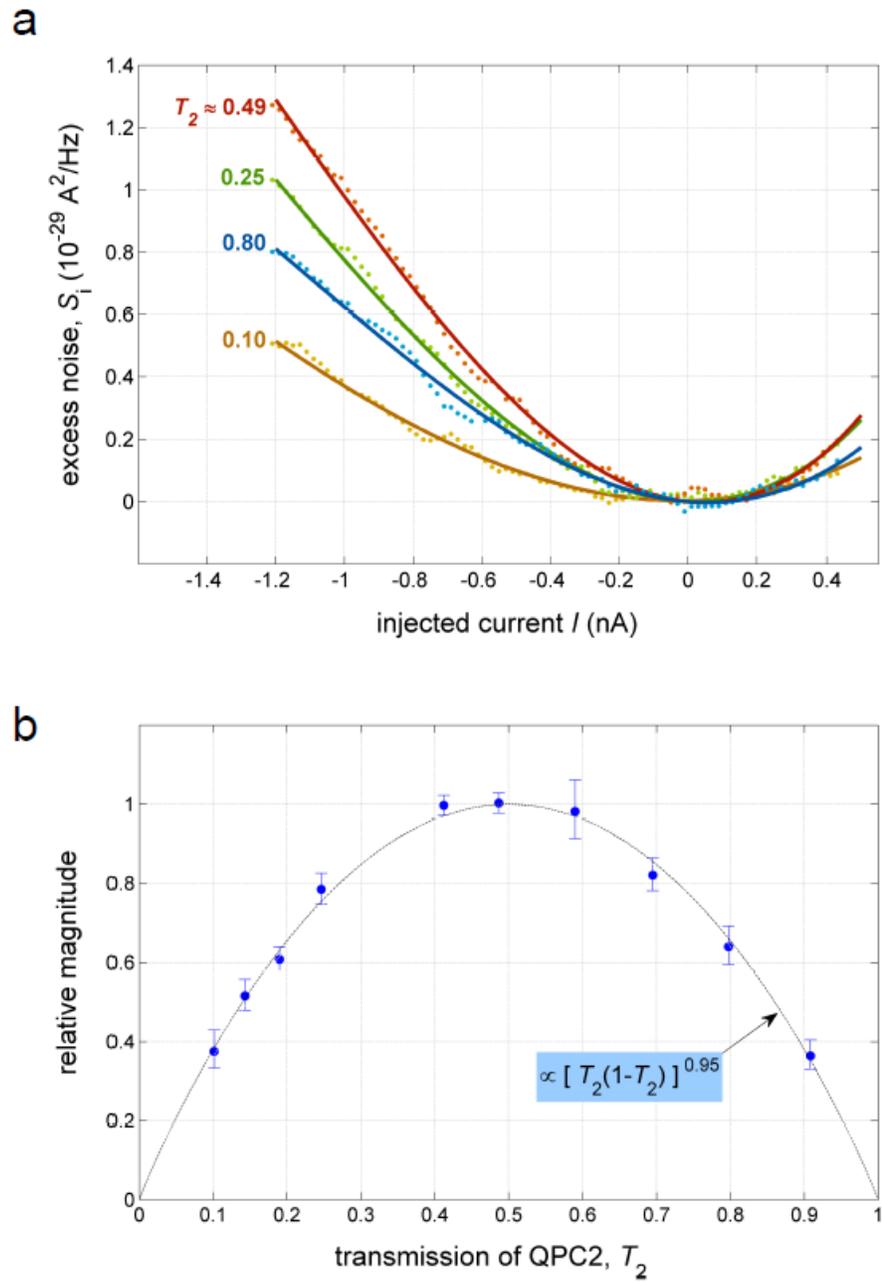

Figure4

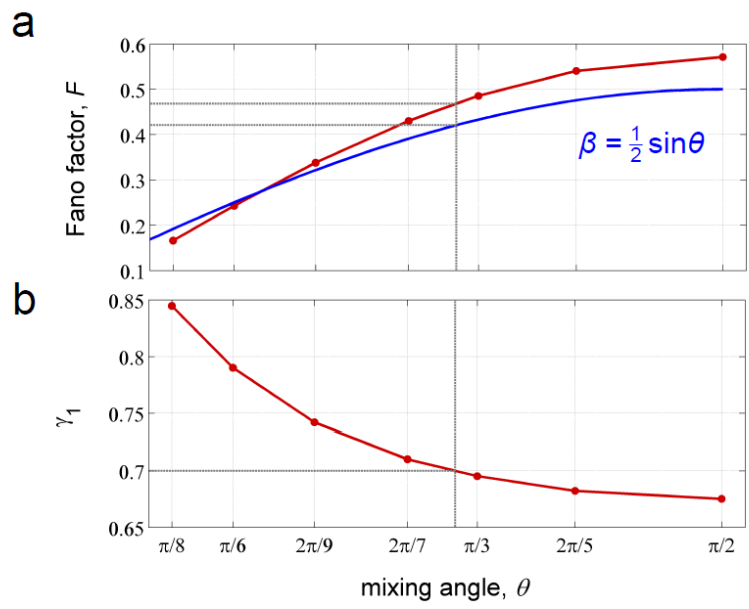